\documentclass[final,3p,times]{elsarticle}
\usepackage{graphics,epsfig,subfigure}
\usepackage{amssymb,amsmath}
\usepackage{cases}
\usepackage{verbatim}
\usepackage{amstext}
\usepackage{color}
\usepackage{array}
\usepackage{multirow}
\newcommand\beq{\begin{equation}}
\newcommand\eeq{\end{equation}}
\newcommand\beqn{\begin{eqnarray}}
\newcommand\eeqn{\end{eqnarray}}

\biboptions{comma,square,compress,numbers,sort}

\journal{Physics Letters B}

\begin{document}

\begin{frontmatter}

\title{Novel equal area law and analytical charge-electric potential criticality for charged Anti-de Sitter black holes}

\author[label1]{Run Zhou}
 \author[label1,label2]{Shao-Wen Wei\corref{cor1}}
  \ead{weishw@lzu.edu.cn}
  \cortext[cor1]{The corresponding author.}

\address[label1]{Institute of Theoretical Physics $\&$ Research Center of Gravitation, Lanzhou University, Lanzhou 730000, People's Republic of China}
\address[label2]{Department of Physics and Astronomy, University of Waterloo, Waterloo, Ontario, Canada, N2L 3G1}

\begin{abstract}
We study the equal area law and charge-electric potential criticality for the charged Anti-de Sitter black holes. Considering that the black hole charge is a double-valued function of the electric potential, we investigate the equal area law in detail. We find that the equal area law has two different expressions when the thermodynamic quantities are near the critical point and far from the critical point. For these two different cases, we obtain the analytical coexistence curve for the low and high electric potential black hole phases by using these two expressions of the equal area law. Based on the result, we analytically study the phase diagram and the critical phenomena in the charge-electric potential plane.
\end{abstract}


\end{frontmatter}

\section{Introduction}

Since the establishment of the four laws of black hole thermodynamics, phase transition and critical phenomena continue to be an exciting and challenging topic in black hole physics. It was Hawking and Page, who first discovered the existence of phase transition between the stable large black hole and thermal gas in an AdS space \cite{Hawking}. Especially, motivated by the AdS/CFT correspondence \cite{Maldacena,Gubser,Witten}, the Hawking-Page phase transition was explained as the confinement/deconfinement phase transition of the gauge field \cite{Witten2}. Subsequently, the study of the black hole phase transition in AdS space attracts more attention.

Interestingly, there exist stable small and large charged or rotating black hole in AdS space. Among them, there is a phase transition of the van der Waals (vdW) type \cite{Chamblin,Chamblin2,Cognola}. Very recently, the study of the black hole thermodynamics has been further generalized to the extended phase space. In this parameter space, the cosmological constant was interpreted as the thermodynamic pressure \cite{Kastor,Dolan00,Cvetic, Yang18}
\begin{eqnarray}
P=-\frac{\Lambda}{8\pi}=\frac{3}{8\pi l^{2}}.\label{ppll}
\end{eqnarray}
The corresponding conjugate quantity is the thermodynamic volume. Then the small-large black hole phase transition was reconsidered in Ref. \cite{Kubiznak} for the charged AdS black holes. Significantly, the results state that the black hole systems and the vdW fluid have the similar phase structure and $P$-$V$ (pressure-thermodynamic volume) criticality. Therefore, it is natural to conjecture that they may have the similar microstructure \cite{WeiLiu}. The study has been extended to other black hole backgrounds. Besides the small-large black hole phase transition, more interesting phase transitions were found, for review see \cite{Altamirano,Teo}.

Comparing with the $P$-$V$ criticality, the $Q$-$\Phi$ (charge-electric potential) criticality has also attracted much more attention. It allows a low electric potential black hole transits to a high electric potential black hole triggered by the temperature or pressure. The phase diagram and critical phenomena were studied for the different AdS black holes \cite{Wu,Shen,Niu,Tsai,Wei,Ma,Hendi11,Hendi12,Hendi13,Hendi14}.

On the other hand, as an alternative method of Gibbs free energy to determine the phase transition point, the equal area laws were investigated. As early as in \cite{Chamblin2}, the authors started with the first law and showed that there is the equal area law in the $T$-$S$ (temperature-entropy) plane for the charged AdS black holes. Employing this equal area law, the first analytical coexistence curve was obtained for the four dimensional charged AdS black holes \cite{Spallucci}. Based on it, the analytical study of the critical phenomena becomes possible. In Ref. \cite{WeiLiu2}, we started from the first law of the charged AdS black hole, and showed that there exist three kinds of the equal area laws in $T$-$S$, $P$-$V$, and $Q$-$\Phi$ planes. Moreover, for the rotating black holes, the equal area law also holds in $J$-$\Omega$ (angular momentum-angular velocity) plane \cite{Wei1}. This result also clarifies that the equal area law does not valid in the $P$-$v$ (pressure-specific volume) plane. Other work concerning the equal area law can be found in \cite{Belhaj,Lan,Miao,Zhao}.

Although it is shown that the equal area laws must hold in different planes, one needs to construct them in detail. For the four dimensional charged AdS black hole, equal area laws in $T$-$S$, $P$-$V$ planes have been well constructed. However, in the $Q$-$\Phi$ plane, the detailed construction is still not given. So in this paper, we will carry out the investigation of the equal area law in the $Q$-$\Phi$ plane for the charged AdS black hole. When plotting the isothermal and isobaric lines in the $Q$-$\Phi$ plane, we find that the equal area law has quite different behaviors when the thermodynamic parameters are far from the critical point. One is the typical case, while another is a novel one, which is mainly caused by the fact that the black hole charge is a double-valued function of the electric potential. For these two different cases, we, respectively, obtain the expressions of the equal area laws. Based on them, we obtain the analytical coexistence curve in the $Q$-$\Phi$ plane. Moreover, employing the form of the coexistence curve, we analytically study the phase diagram, order parameter, and $Q$-$\Phi$ criticality.

The paper is structured as follows. In Sec. \ref{laww}, we start with the first law, and consider the equal area for a thermodynamic system. Then considering the chemical potential term is a double-valued function, we obtain two different expressions for the equal area law. Since the black hole charge $Q$ is a double-valued function of the electric potential $\Phi$, we generalize the study to the charged AdS black hole in Sec. \ref{null}. Employing these two different expressions of the equal area law, we obtain the analytical coexistence curve in the $Q$-$\Phi$ plane. In Sec. \ref{exponent}, we analytically study the phase diagram, order parameter, and the critical exponent. Finally, a brief summary is given in Sec. \ref{Summary}.

\section{First law and equal area laws}
\label{laww}

Before pursuing the specific equal are law in the $Q$-$\Phi$ plane for the charged AdS black hole, we would like to investigate the equal area law for an arbitrary thermodynamic system.

Let us start with the first law
\begin{eqnarray}
dE=TdS-PdV+xdy.
\end{eqnarray}
The last term $xdy$ denotes the chemical potential term. The Gibbs free energy $G=E-PV-TS$, which leads to
\begin{eqnarray}
dG=-SdT+VdP+xdy.
\end{eqnarray}
In fact, we can absorb the -$SdT$ and $VdP$ terms into the third term, i.e., 
\begin{eqnarray}
dG=\sum_{i}x^{i}dy^{i}.\label{iii}
\end{eqnarray}
For simplicity, we only let one $y^{i}$ vary freely, while keep others fixed, i.e., $dG=xdy$. Considering that the system undergoes a first order phase transition at constant $y^{*}$ from $x_{1}$ to $x_{3}$, where the system phases located at $x_{1}$ and $x_{3}$ are two coexistence phases, one has $\Delta G=0$. Therefore, integrating (\ref{iii}),  we can express it in a circular integral form
\begin{eqnarray}
 \oint_{c} xdy=0.\label{ealss}
\end{eqnarray}
Here the circular integral path is $x_{1}(y^{*})\stackrel{x(y)}{\longrightarrow} x_{3}(y^{*})\stackrel{y=y^{*}}{\longrightarrow} x_{1}(y^{*})$. This formula (\ref{ealss}) is just the expression of the Maxwell equal area law. Employing it, we can exactly determine the value of $y^{*}$. If taking $y=P$ and $x=V$, we will have the equal area law in the $P$-$V$ plane, i.e., $\oint_{c}VdP=0$. In general, it is more convenient to express $y$ in terms of $x$ rather than express $x$ in terms of $y$. So next, we will change the integral variable $y$ to $x$ in Eq. (\ref{ealss}). However, for some cases, $y$ has double values for each $x$, which makes the situation much more interesting. Since it is not noted before, we will examine it in detail in the following section.

\begin{figure}
	\center{\subfigure[]{\label{yx1}
			\includegraphics[width=7cm]{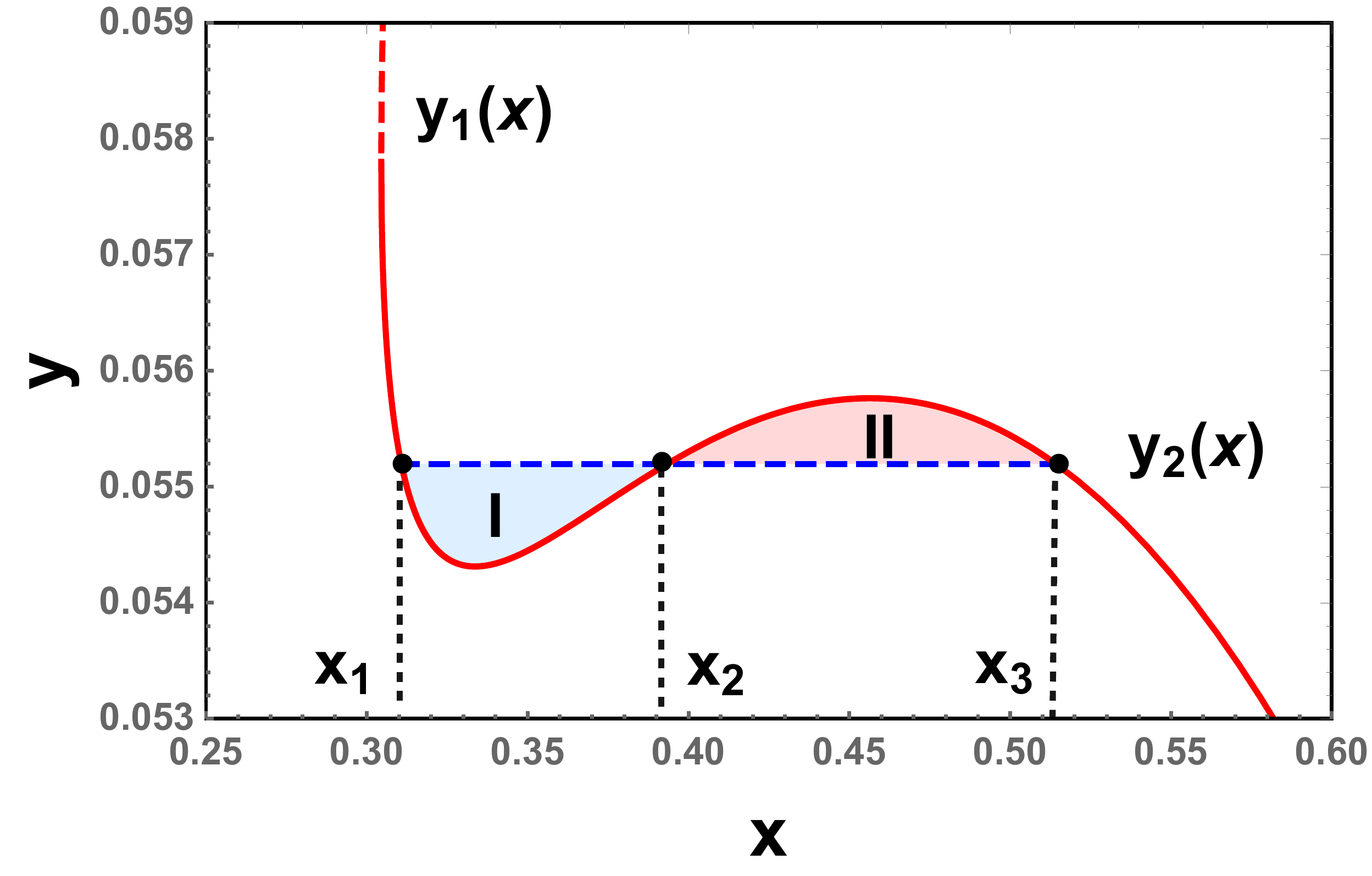}}
		\subfigure[]{\label{yx2}
			\includegraphics[width=7cm]{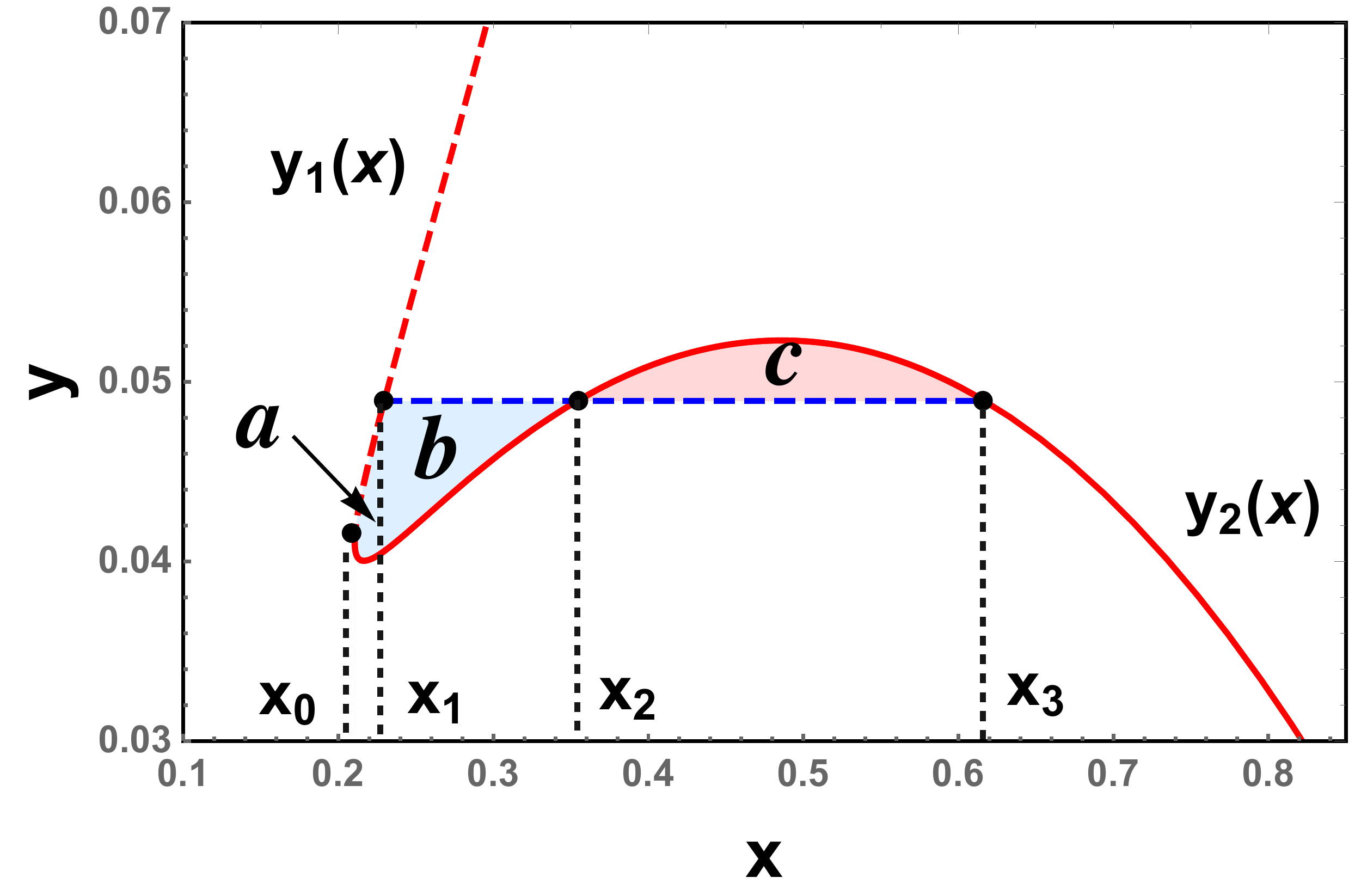}}}
\caption{Equal area law in the $y$-$x$ plane. $y$ is a double function of $x$, while functions $y_{1}(x)$ (red dashed lines) and $y_{2}(x)$ (red solid lines) are single-valued functions of $x$. (a) Only $y_{2}(x)$ participates in the construction of the equal area law. The horizontal line $y=y^{*}$ intersects $y_{2}(x)$ at $x_{1}$, $x_{2}$, and $x_{3}$ with $x_{1}<x_{2}<x_{3}$. (b) Both $y_{1}(x)$ and $y_{2}(x)$ are used for the equal area law. The horizontal line $y=y^{*}$ intersects $y_{1}(x)$ at $x_{1}$, and $y_{2}(x)$ at $x_{2}$ and $x_{3}$. Functions $y_{1}(x)$ and $y_{2}(x)$ connect at $x_{0}$. The relation $x_{0}<x_{1}<x_{2}<x_{3}$ holds.}\label{ppyx1}
\end{figure}

\subsection{Case I}
\label{ca1}

Here we consider that $y$ is a double-valued function of $x$, while it can be separated into two single-valued functions, $y_{1}(x)$ and $y_{2}(x)$. The equal area law is constructed by only one branch of $y$, see Fig. \ref{yx1}. Actually, this case is like that $y$ is a single-valued function of $x$. The horizontal line $y=y^{*}$ intersects $y_{2}(x)$ (red solid curve) at $x_{1}$, $x_{2}$, and $x_{3}$ with $x_{1}<x_{2}<x_{3}$.

Typically, the equal area law (\ref{ealss}) can be easily re-expressed as
\begin{eqnarray}
 \oint_{c} y_{2}dx=\int_{x_{1}}^{x_{3}} y_{2}dx+\int_{x_{3}}^{x_{1}} y^{*}dx=0.
\end{eqnarray}
Since $y^{*}$ is a constant, it reduces to
\begin{eqnarray}
 \int_{x_{1}}^{x_{3}} y_{2}dx=y^{*}(x_{3}-x_{1}).\label{eals}
\end{eqnarray}
This is a typical form that we usually used. Making use this formula, we can determine the value of $y^{*}$ of the phase transition point.

\subsection{Case II}
\label{ca2}

In this case, both single-valued functions $y_{1}(x)$ and $y_{2}(x)$ are included in constructing the equal area law. We plot a sketch picture for them in Fig. \ref{yx2}. Functions $y_{1}(x)$ and $y_{2}(x)$ are plotted in red dashed line and red solid line, respectively. The horizontal line $y=y^{*}$ intersects $y_{1}(x)$ at $x_{1}$, and intersects $y_{2}(x)$ at $x_{2}$ and $x_{3}$. Functions $y_{1}(x)$ and $y_{2}(x)$ connect at $x_{0}$, while which is not the extremal point of $y_{2}(x)$. Moreover, the relation $x_{0}<x_{1}<x_{2}<x_{3}$ holds.

These three areas in the figure marked in shadow can be calculated as
\begin{eqnarray}
{\rm area}(a)&=&-\int_{x_{1}}^{x_0}y_1(x)dx-\int_{x_0}^{x_{1}}y_2(x)dx,\label{a1}\\
{\rm area}(b)&=&-\int_{x_{1}}^{x_{2}} y_2(x)dx+y^*(x_{2}-x_{1}),\\
{\rm area}(c)&=&\int_{x_{2}}^{x_{1}} y_2(x)dx-y^*(x_{3}-x_{2}).\label{a3}
\end{eqnarray}
If $y^{*}$ is the phase transition point and the equal area law (\ref{ealss}) holds, one must have
\begin{eqnarray}
{\rm area}(a)+{\rm area}(b)={\rm area}(c). \label{a4}
\end{eqnarray}
Plugging (\ref{a1})-(\ref{a3}) into (\ref{a4}), we obtain
\begin{equation}
 \int_{x_{1}}^{x_0}y_1(x)dx+\int_{x_0}^{x_{3}}y_2(x)dx=y^*(x_{3}-x_{1}).\label{eqa2}
\end{equation}
Obviously, this novel expression of the equal area law is different from the typical one given in (\ref{eals}). Thus, one must be very careful when studying the phase transition for a thermodynamic system by using the equal area law.

\section{Equal area law for charged AdS black hole}
\label{null}

Now, it is widely known that there exists a small-large black hole phase transition in the background of charged AdS black hole. Here we would like to investigate its coexistence curve by using the equal area law in the $Q$-$\Phi$ plane. 

The line element to describe this charged AdS black hole is
\begin{eqnarray}
ds^{2}=-f(r)dt^{2}+\frac{dr^{2}}{f(r)}+r^{2}(d\theta^{2}+\sin^{2}\theta d\phi^{2}),
\end{eqnarray}
where the function is given by
\begin{eqnarray}
f(r)=1-\frac{2M}{r}+\frac{Q^{2}}{r^{2}}+\frac{r^{2}}{l^{2}}.
\end{eqnarray}
Here the parameter $M$ and $Q$ are the black hole mass and charge, respectively. The AdS radius $l$ is related to the pressure by Eq. (\ref{ppll}). The temperature $T$, entropy $S$, electric potential $\Phi$, thermodynamic volume $V$, and Gibbs free energy $G$ are 
\begin{eqnarray}
T&=&2Pr_{\rm h}+\frac{1}{4\pi r_{\rm h}}-\frac{Q^{2}}{4\pi r_{\rm h}^{3}},\quad
S=\pi r_{\rm h}^{2},\quad
\Phi=\frac{Q}{r_{\rm h}},\\
V&=&\frac{4}{3}\pi r_{\rm h}^{3},\quad
G=\frac{r_{\rm h}}{4}-\frac{2\pi P r_{\rm h}^{3}}{3}+\frac{3Q^{2}}{4r_{\rm h}},
\end{eqnarray}
with $r_{\rm h}$ being the horizon radius of the black hole. Solving the pressure from the equation of the temperature, one can get the state equation for the black hole, which reads
\begin{eqnarray}
P=\frac{T}{v}-\frac{1}{2\pi v^{2}}+\frac{2Q^{2}}{\pi v^{4}},\label{steq}
\end{eqnarray}
where $v=2r_{\rm h}$ is the specific volume of the black hole fluid. Using the condition $(\partial_{v}P)=(\partial_{v,v}P)=0$, one can easily obtain the critical point \cite{Kubiznak}
\begin{eqnarray}
T_{\rm c}=\frac{\sqrt{6}}{18\pi Q},\quad
v_{\rm c}=2\sqrt{6}Q,\quad
P_{\rm c}=\frac{1}{96\pi Q^{2}}.\label{cp}
\end{eqnarray}
This point corresponds to a second order phase transition. Below the point, the system will encounter a first order phase transition. The isothermal and isobaric lines also show the non-monotonous behaviors with $v$ or $r_{\rm h}$. More interestingly, the Gibbs free energy demonstrates the swallow tail behavior.

\begin{figure}
	\center{\subfigure[]{\label{Tbhs}
		\includegraphics[width=7cm]{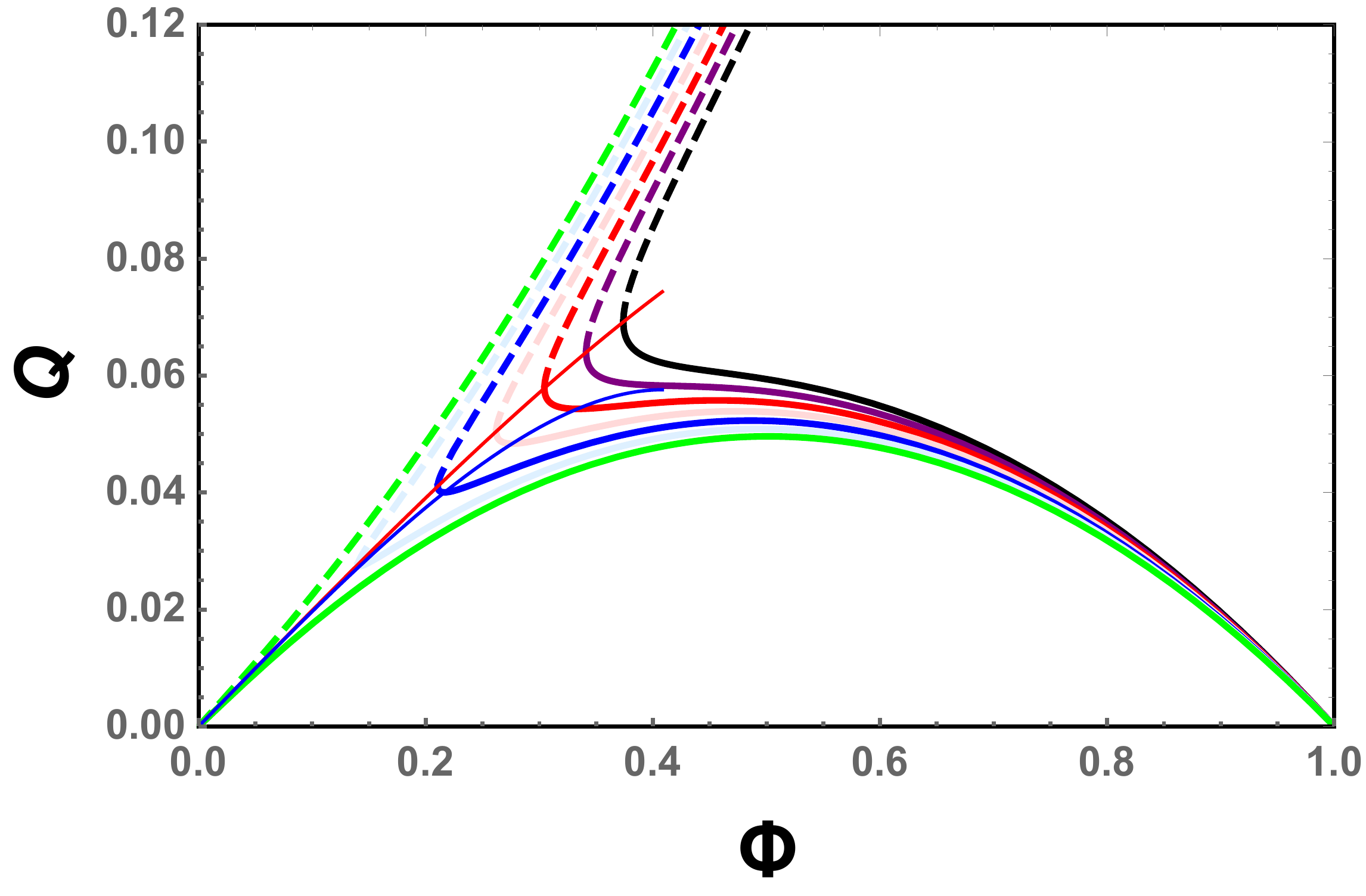}}
		\subfigure[]{\label{TQPtwoc}
		\includegraphics[width=7cm]{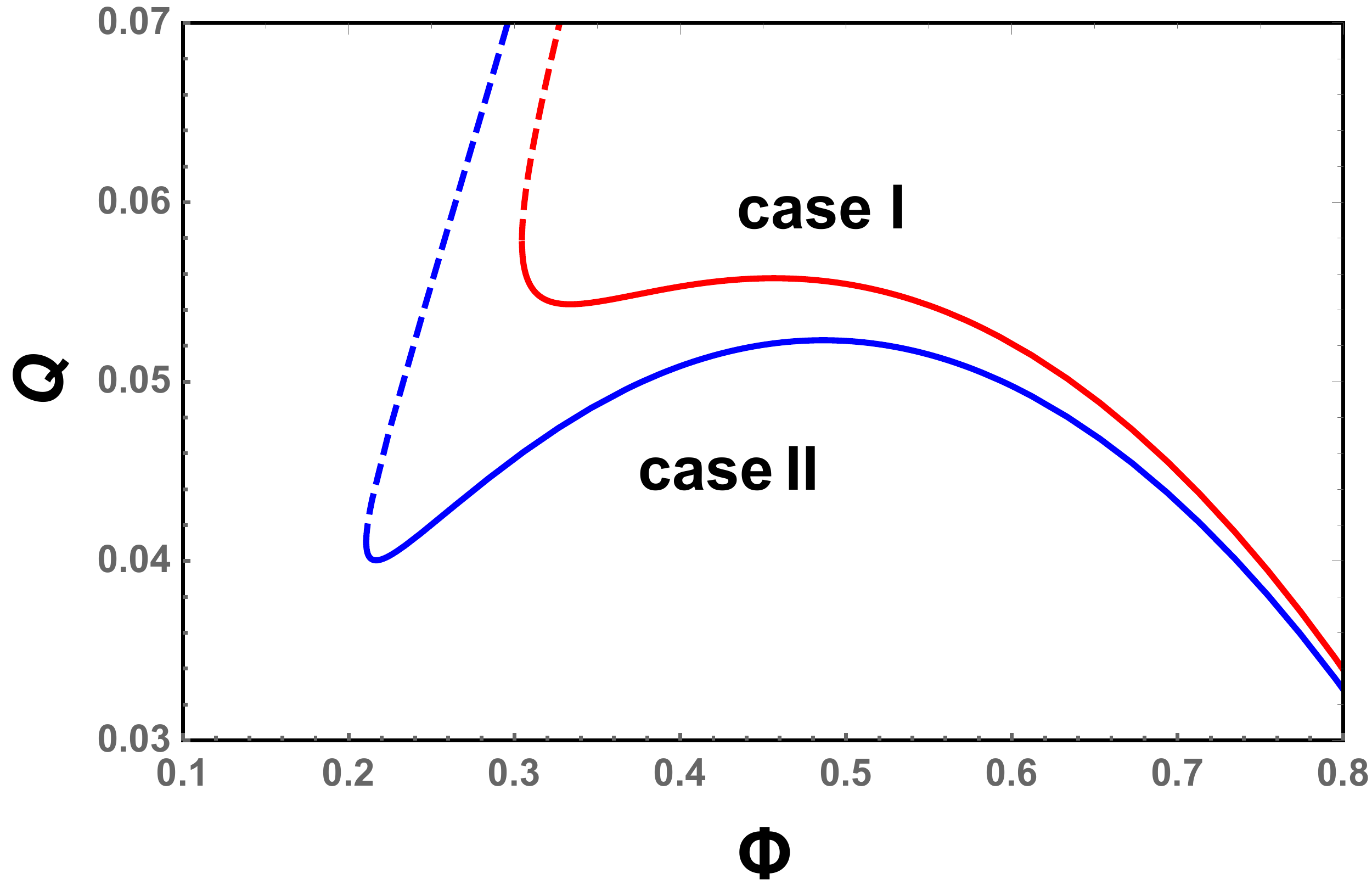}}}
	\caption{(a) The behavior of the charge $Q$ as a function of the electric potential $\Phi$ with $P$=1. The temperature $T$ varies from 0.74 to 0.80 from top to bottom. The dashed and solid lines are for $Q_{1}$ and $Q_{2}$, respectively. The thin red line denotes the connection point $Q_{0}$ of the curves $Q_{1}$ and $Q_{2}$ and the thin blue line is for the minimum point $Q_{2}^{\rm m}$.  (b) Two different behaviors of charge $Q_{2}$. Case I is very near the critical point, and case II is far from the critical point.}\label{ppTNss}
\end{figure}

Employing the expression of these thermodynamic quantities, the state equation (\ref{steq}) can be expressed as
\begin{eqnarray}
\Phi^{4}-\Phi^{2}+4\pi TQ\Phi-8\pi P Q^{2}=0.
\end{eqnarray}
Solving $Q$ from the equation, we have two solutions
\begin{eqnarray}
Q_{1}&=&\frac{\sqrt{\pi}T+\sqrt{2P\Phi^{2}+\pi T^{2}-2P}}{4\sqrt{\pi} P}\Phi,\\
Q_{2}&=&\frac{\sqrt{\pi}T-\sqrt{2P\Phi^{2}+\pi T^{2}-2P}}{4\sqrt{\pi} P}\Phi.\label{qqq2}
\end{eqnarray}
So the charge $Q$ is a double-valued function of $\Phi$, while $Q_{1}$ and $Q_{2}$ are single-valued functions. Moreover, $Q_{1}$ and $Q_{2}$ meet each other at
\begin{eqnarray}
Q_{0}=\frac{\Phi\sqrt{1-\Phi^{2}}}{2\sqrt{2\pi P}}.\label{q00}
\end{eqnarray}
A simple calculation show that $Q_{1}$ has no extremal point. While $Q_{2}$ has one minimum and one maximum, and both of them share the same expression
\begin{eqnarray}
Q_{2}^{\rm m}=\frac{\Phi\sqrt{1-3\Phi^{2}}}{2\sqrt{2\pi P}}.
\end{eqnarray}
In Fig. \ref{Tbhs}, we plot the charge $Q$ as a function of $\Phi$ with fixing $P$ and $T$. The pressure $P=1$, and the temperature $T$ varies from 0.74 to 0.8 from top to bottom. From the critical point (\ref{cp}), one can find that the critical temperature corresponded to $P=1$ is $T_{\rm c}=0.7523$. It is obvious that $Q_{1}$ always increases with $\Phi$. While for $Q_{2}$, it has a different behavior. If the temperature $T<T_{\rm c}$, $Q_{2}$ decreases with $\Phi$. However, when $\sqrt{\frac{2P}{\pi}}>T>T_{\rm c}$, $Q_{2}$ firstly decreases, then increases, and finally decreases with $\Phi$. While when $T>\sqrt{\frac{2P}{\pi}}$,  $Q_{2}$ firstly increases, then decreases with $\Phi$.

Next, we would like to determine the phase transition point by constructing the equal area laws in the $Q$-$\Phi$ plane. In Sec. \ref{laww}, we have discussed the area law for a thermodynamic system when $y$ is a double-valued function of x, so we can do the following replacement,
\begin{eqnarray}
x\rightarrow \Phi, \quad y_{1}\rightarrow Q_{1}, \quad y_{2}\rightarrow Q_{2}.\label{rep}
\end{eqnarray}
Considering the behaviors of the charge $Q$, we can divide them into two cases, see Fig. \ref{TQPtwoc}. Case I is very near the critical point, and case II is far from the critical point. In the first case, the constant phase transition charge $Q^{*}$ only intersects with $Q_{2}$. For the second case, the constant phase transition charge intersects $Q_{1}$ at the left and intersects $Q_{2}$ at the right.  

\subsection{Case I: near the critical point}

For this case described in Fig. \ref{TQPtwoc}, we can find that only the solution $Q_{2}$ is enough to construct these two equal areas. This is consistent with that given in Sec. \ref{ca1}, so the equal are law can be expressed as
\begin{eqnarray}
 \int_{\Phi_{1}}^{\Phi_{3}} Q_{2}dx=Q^{*}(\Phi_{3}-\Phi_{1}).\label{eqa}
\end{eqnarray}
Plunging $Q_{2}$ given in (\ref{qqq2}) into it, we have
\begin{eqnarray}
Q^*=\frac{1}{(\Phi_{3}-\Phi_{1})}\bigg(\frac{T}{8P}(\Phi_{3}^{2}-\Phi_{1}^{2})+\frac{\left(\pi T^{2}+2P(-1+\Phi_{1}^{2})\right)^{\frac{3}{2}}}{24\sqrt{\pi}P^{2}}-\frac{\left(\pi T^{2}+2P(-1+\Phi_{3}^{2})\right)^{\frac{3}{2}}}{24\sqrt{\pi}P^{2}}\bigg).\label{pq0}
\end{eqnarray}
On the other hand, from the state equation, we have
\begin{eqnarray}
Q^*&=&\frac{\pi T\Phi_{1}-\sqrt{\pi}\sqrt{-2P\Phi_{1}^{2}+\pi T^{2}\Phi_{1}^{2}+2P\Phi_{1}^{4}}}{4P\pi},\label{pq1}\\
Q^*&=&\frac{\pi T\Phi_{3}-\sqrt{\pi}\sqrt{-2P\Phi_{3}^{2}+\pi T^{2}\Phi_{3}^{2}+2P\Phi_{3}^{4}}}{4P\pi}.\label{pq2}
\end{eqnarray}
Then solving the pressure $P$ from $2\ast(\ref{pq0})=(\ref{pq1})+(\ref{pq2})$, we will get
\begin{equation}
P=\frac{3\pi T^{2}}{2(3+\Phi_{3}^{2}-4\Phi_{3}\Phi_{1}+\Phi_{1}^{2})}.\label{pp1}
\end{equation}
Further, we change (\ref{pq1}) and (\ref{pq2}) into the following forms
\begin{eqnarray}
P&=&\frac{\Phi_{1}(4\pi Q^* T-\Phi_{1}+\Phi_{1}^{3})}{8\pi {Q^*}^{2}},\label{pp2}\\
P&=&\frac{\Phi_{3}(4\pi Q^* T-\Phi_{3}+\Phi_{3}^{3})}{8\pi {Q^*}^{2}}.\label{pp3}
\end{eqnarray}
Thus, by solving Eqs. (\ref{pp1}), (\ref{pp2}), and (\ref{pp3}), one can obtain the coexistence curve in the $Q$-$\Phi$ plane. Taking (\ref{pp2})-(\ref{pp3})=0 and $2\ast(\ref{pp1})=(\ref{pp2})+(\ref{pp3})$, we arrive
\begin{align}
&-(\Phi_{3}+\Phi_{1})+(\Phi_{3}+\Phi_{1})\left(-2\Phi_{3}\Phi_{1}+(\Phi_{3}+\Phi_{1})^{2}\right)+4\pi Q^* T=0,\\
&\frac{2\Phi_{3}\Phi_{1}-2\Phi_{3}^{2}\Phi_{1}^{2}-(\Phi_{3}+\Phi_{1})^{2}+\left(-2\Phi_{3}\Phi_{1}+(\Phi_{3}+\Phi_{1})^{2}\right)^{2}+4(\Phi_{3}+\Phi_{1})\pi Q^* T}{8\pi {Q^*}^{2}}-\frac{3\pi T^{2}}{3-6\Phi_{3}\Phi_{1}+(\Phi_{3}+\Phi_{1})^{2}}=0.
\end{align}
Solving these equations, we can express $\Phi_{1}$ and $\Phi_{3}$ in terms of the charge $Q$ and temperature $T$
\begin{eqnarray}
\Phi_{1}&=&\left(\frac{1}{24A}\right)^{\frac{1}{3}}+\left(\frac{A}{72}\right)^{\frac{1}{3}}-\frac{B}{6A^{\frac{1}{3}}},\label{ww0}\\
\Phi_{3}&=&\left(\frac{1}{24A}\right)^{\frac{1}{3}}+\left(\frac{A}{72}\right)^{\frac{1}{3}}+\frac{B}{6A^{\frac{1}{3}}},
\end{eqnarray}
where
\begin{eqnarray}
A&=&-9\pi Q^* T+\sqrt{81\pi^{2} {Q^*}^{2} T^{2}-3},\\
B&=&\sqrt{\frac{3\times 3^{2/3}+9\times (3A^{2})^{1/3}+9A^{4/3}+3^{2/3}A^{2}-36\times3^{2/3}A\pi Q^*T}{3^{1/3}+A^{2/3}}}.
\end{eqnarray}
Plugging $\Phi_{1}$ and $\Phi_{3}$ into (\ref{pp1}), one can obtain a relation between the pressure, temperature and charge \cite{Spallucci} 
\begin{eqnarray}
T^{2}=\frac{8P(3-\sqrt{96\pi P {Q^*}^{2}})}{9\pi}.
\end{eqnarray}
Making using this equation, $\Phi_{1}$ and $\Phi_{3}$ can be further expressed as
\begin{eqnarray}
\Phi_{1}&=&\sqrt{\frac{3-8Q^*\sqrt{6\pi P}-\sqrt{9-48Q^*\sqrt{6\pi P}+288 \pi {Q^*}^{2}P}}{6}},\label{pps}\\
\Phi_{3}&=&\sqrt{\frac{3-8Q^*\sqrt{6\pi P}+\sqrt{9-48Q^*\sqrt{6\pi P}+288 \pi {Q^*}^{2}P}}{6}}.\label{ppl}
\end{eqnarray}
Moreover, we can solve the charge of the phase transition point from the equation, which gives
\begin{eqnarray}
Q^*=\frac{\sqrt{3}\Phi_{1,3}(\sqrt{3\Phi_{1, 3}^{2}+1}-2\Phi_{1, 3})}{2\sqrt{2\pi P}}.\label{ww1}
\end{eqnarray}
Actually, this describes the coexistence curve of the phase transition, and the phase diagram can be well determined by it. However, we need to note that this equation only effective for the black hole system very near the critical case. When the thermodynamic quantities are far from the critical point, we need to consider the second case.

\subsection{Case II: far from the critical point}

For this case, we can find that both $Q_{1}$ and $Q_{2}$ are useful in constructing the equal area laws. The charge $Q^{*}$ of the phase transition point will intersect $Q_{1}$ at $\Phi_{1}$, and intersect $Q_{2}$ at $\Phi_{2}$ and $\Phi_{3}$. Thus this case is exactly consistent with that given in Sec. \ref{ca2}.

Taking the replacement (\ref{rep}), the equal area law (\ref{eqa2}) reduces to
\begin{equation}
Q^*(\Phi_{3}-\Phi_{1})=\int_{\Phi_{1}}^{\Phi_0}Q_1(\Phi)d\Phi+\int_{\Phi_0}^{\Phi_{3}}Q_2(\Phi)d\Phi.\label{jfs}
\end{equation}
It is worthwhile pointing out that $\Phi_{1}>\Phi_0$, so the first term in the right side is negative. Plugging $Q_{1}$ and $Q_{2}$ into (\ref{jfs}), we integrate it and get
\begin{equation}
Q^*(\Phi_{3}-\Phi_{1})=\frac{T}{8P}(\Phi_{3}^{2}-\Phi_{1}^{2})-\frac{\left(\pi T^{2}+2P(-1+\Phi_{1}^{2})\right)^{\frac{3}{2}}}{24\sqrt{\pi}P^{2}}-\frac{\left(\pi T^{2}+2P(-1+\Phi_{3}^{2})\right)^{\frac{3}{2}}}{24\sqrt{\pi}P^{2}},\label{dd0}
\end{equation}
where we have used $\Phi_0=\sqrt{\frac{2P-\pi T^2}{2P}}$ corresponding to the connection point of $Q_{1}$ and $Q_{2}$ . At the phase transition points $\Phi_{1}$ and $\Phi_{3}$, they, respectively, satisfy 
\begin{eqnarray}
Q^*&=&\frac{\pi T\Phi_{1}+\sqrt{\pi}\sqrt{-2P\Phi_{1}^{2}+\pi T^{2}\Phi_{1}^{2}+2P\Phi_{1}^{4}}}{4P\pi},\label{dd2}\\
Q^*&=&\frac{\pi T\Phi_{3}-\sqrt{\pi}\sqrt{-2P\Phi_{3}^{2}+\pi T^{2}\Phi_{3}^{2}+2P\Phi_{3}^{4}}}{4P\pi}.\label{dd1}
\end{eqnarray}
Making use (\ref{dd0})-(\ref{dd1}), we have
\begin{eqnarray}
&&3P(\Phi_{3}-\Phi_{1})\left (\Phi_{1}\sqrt{-2P+\pi T^{2}+2P\Phi_{1}^{2}}-\Phi_{3}\sqrt{-2P+\pi T^{2}+2P\Phi_{3}^{2}}\right)\nonumber\\
&&=-(-2P+\pi T^{2}+2P\Phi_{1}^{2})^{\frac{3}{2}}-(-2P+\pi T^{2}+2P\Phi_{3}^{2})^{\frac{3}{2}}.
\end{eqnarray}
Solving the pressure from it, one can obtain
\begin{equation}
P=\frac{3\pi T^{2}}{2(3+\Phi_{3}^{2}-4\Phi_{3}\Phi_{1}+\Phi_{1}^{2})}.
\end{equation}
Interestingly, this pressure is exactly the same as (\ref{pp1}) of case I. Moreover, from (\ref{dd2}) and (\ref{dd1}), we arrive
\begin{eqnarray}
P&=&\frac{\Phi_{1}(4\pi Q^* T-\Phi_{1}+\Phi_{1}^{3})}{8\pi {Q^*}^{2}},\\
P&=&\frac{\Phi_{\rm l}(4\pi Q^* T-\Phi_{3}+\Phi_{3}^{3})}{8\pi {Q^*}^{2}},
\end{eqnarray}
which is also the same as (\ref{pp2}) and (\ref{pp3}). Thus, it is clear that the equations determining the coexistence curve are exactly the same for both the cases, so the results (\ref{ww0})-(\ref{ww1}) for the first case also valid for the second case.

In summary, according to the equal area laws in the $Q$-$\Phi$ plane, we obtain the coexistence curve, see (\ref{pps})-(\ref{ww1}). Moreover, although the equal area laws have two different expressions near the critical point and far from the critical point, they admit the same coexistence curve. 

\section{Phase diagram and critical exponent}
\label{exponent}

As shown above, the coexistence curve (\ref{ww1}) is effective for the thermodynamic quantities both near and far from their critical values. Based on the result, we in this section would like to study the phase diagram and critical exponent in the $Q$-$\Phi$ plane.

\begin{figure}
	\center{\subfigure[]{\label{TPhasd}
			\includegraphics[width=7cm]{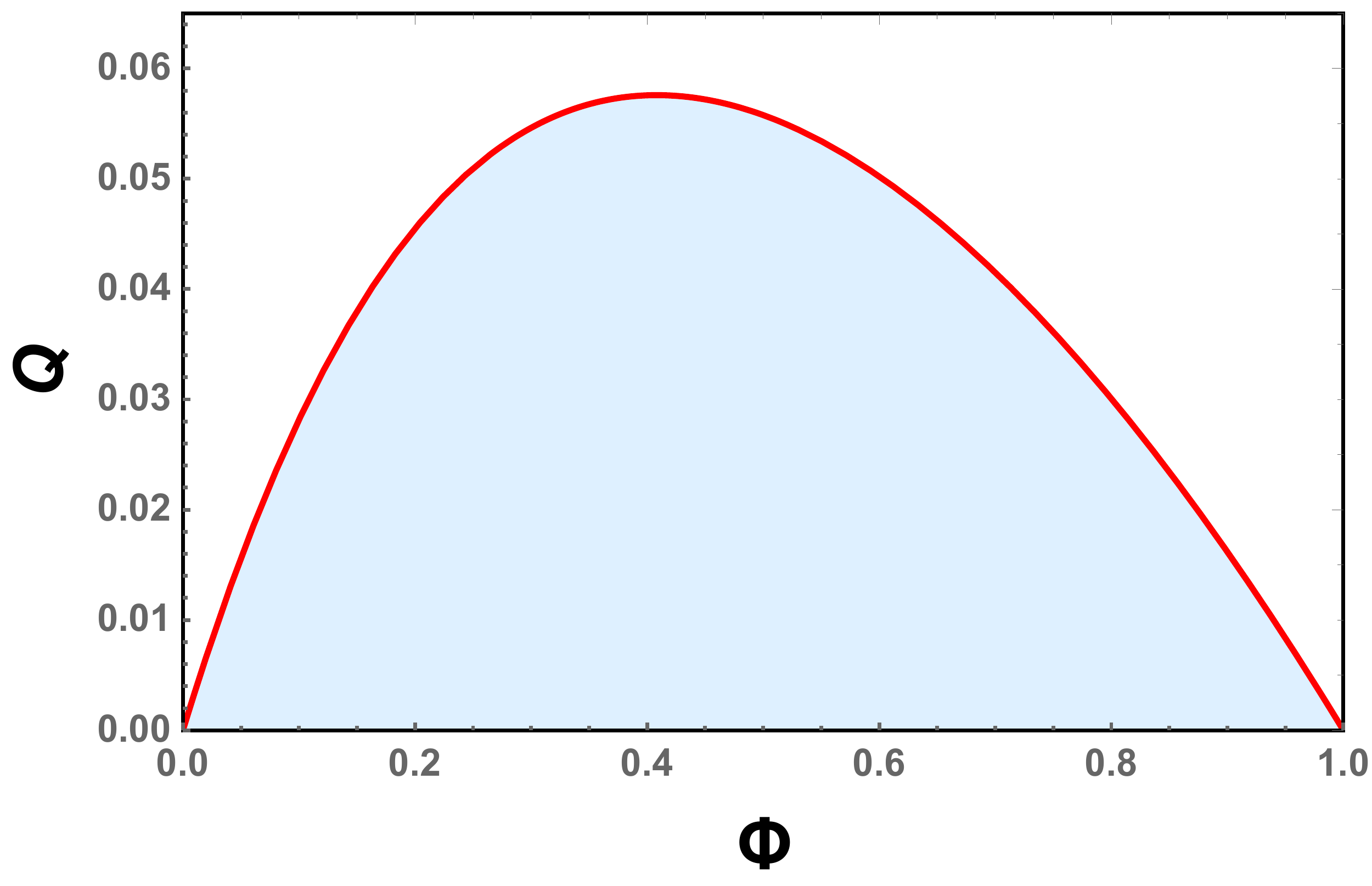}}
		\subfigure[]{\label{TOrderP}
			\includegraphics[width=7cm]{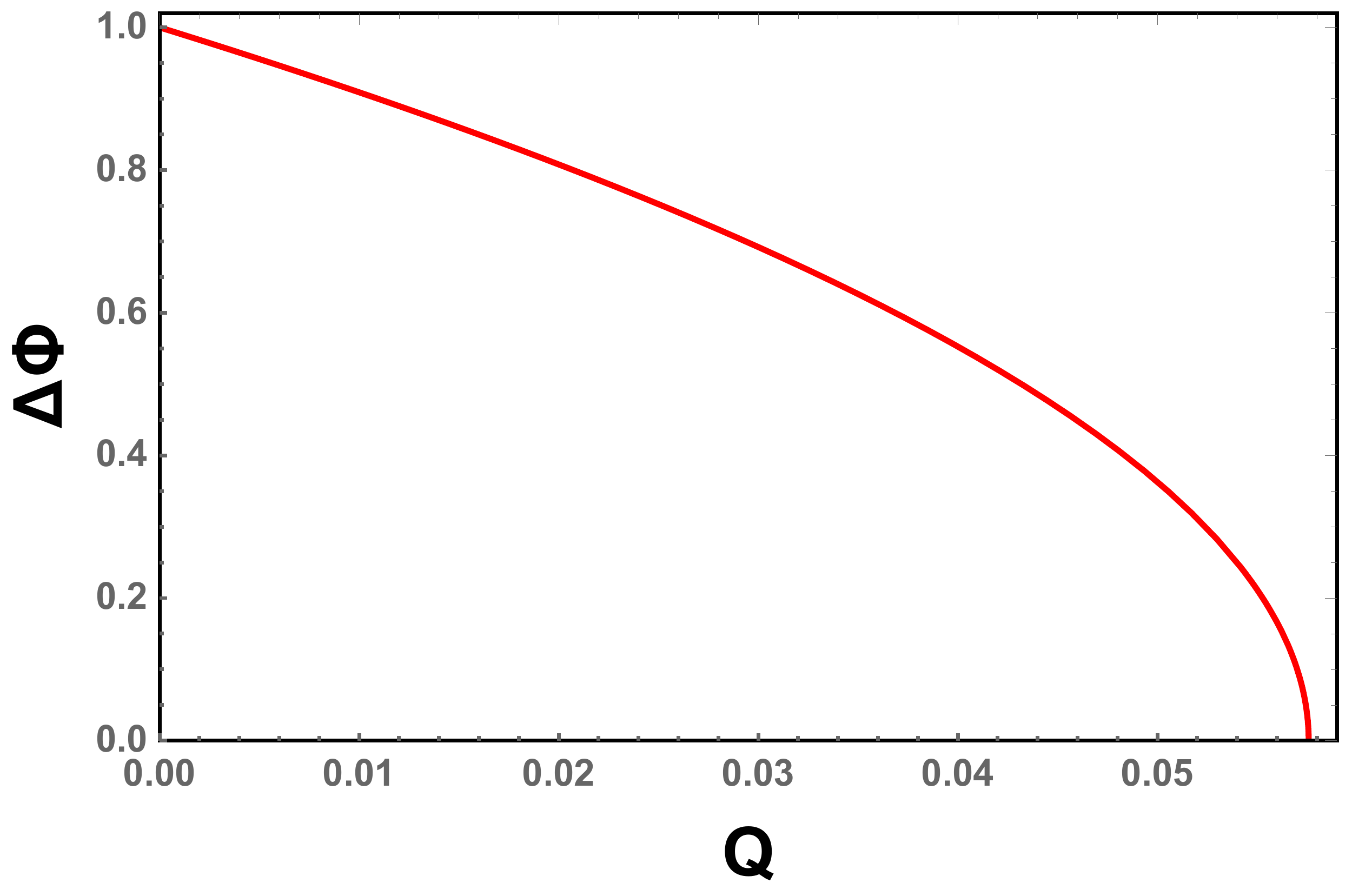}}}
	\caption{(a) Phase diagram in the $Q$-$\Phi$ plane. The shadow region denotes the coexistence phases of the low and high electric potential black holes. (b) Behavior of $\Delta\Phi=\Phi_{3}-\Phi_{1}$ as a function of the charge $Q$. The pressure $P=1$.}\label{ppOrderP}
\end{figure}

First, we plot the coexistence curve in the $Q$-$\Phi$ plane in Fig.  \ref{TPhasd}. To avoiding the confusion, here we name the low-high electric potential black hole phase transition rather than the small-large black hole phase transition. The light blue shadow region denotes the coexistence phase of the low and high electric potential black holes. The low and high electric potential black hole phases are located in the left and right of the figure, respectively. A simple calculation also shows that the boundaries of the coexistence low and high electric potential black hole phases are at $\Phi$=0 and 1, which is independent of the pressure of the black hole system. The critical point is the top point in the coexistence curve. Solving it, we have
\begin{equation}
\Phi_{\rm c}=\frac{1}{\sqrt{6}},\quad Q_{\rm c}=\frac{1}{4\sqrt{6\pi P}}.
\end{equation}
It is clear that $\Phi_{\rm c}$ is a constant and independent of the pressure $P$. While the critical charge $Q_{\rm c}$ decreases with the pressure.

Moreover, we also show the behavior of $\Delta\Phi=\Phi_{3}-\Phi_{1}$ as a function of the charge $Q$ in Fig. \ref{TOrderP}. At $Q=0$, $\Delta\Phi$ has maximum value 1. Then $\Delta\Phi$ decreases with the charge, and approaches zero when the critical charge is achieved. So $\Delta\Phi$ acts as an order parameter, which can be used to describe the low-high electric potential black hole phase transition.

Next, we would like to examine the critical exponents. Since $\Phi_{1}$ and $\Phi_{3}$ have analytical forms, see (\ref{pps}) and (\ref{ppl}), we can expand them near the critical point, the results are
\begin{eqnarray}
(\Phi_{1}-\Phi_{\rm c})&=&-(6\pi P)^{\frac{1}{4}} \left(Q_{\rm c}-Q\right)^{\frac{1}{2}}+\sqrt{\pi P}\left(Q_{\rm c}-Q\right)-\sqrt{\frac{3}{2}} \pi P \left(Q_{\rm c}-Q\right)^{2}+\mathcal{O}\left(Q_{\rm c}-Q\right)^{3},\\
(\Phi_{3}-\Phi_{\rm c})&=&(6\pi P)^{\frac{1}{4}} \left(Q_{\rm c}-Q\right)^{\frac{1}{2}}+\sqrt{\pi P}\left(Q_{\rm c}-Q\right)-\sqrt{\frac{3}{2}} \pi P \left(Q_{\rm c}-Q\right)^{2}+\mathcal{O}\left(Q_{\rm c}-Q\right)^{3}.
\end{eqnarray}
So near the critical point, $(\Phi_{1}-\Phi_{\rm c})$ and $(\Phi_{3}-\Phi_{\rm c})$ have the same exponent of $\frac{1}{2}$. Interestingly, one can find that the first term of these expansions has opposite signs. While other coefficients are exactly the same. Therefore, $\Delta\Phi$ must have the same exponent $\frac{1}{2}$. Actually, with a simple calculation, we have 
\begin{eqnarray}
\Delta\Phi=2(6\pi P)^{\frac{1}{4}} \times\left(Q_{\rm c}-Q\right)^{\frac{1}{2}}.
\end{eqnarray}
Obviously, this confirms the exponent $\frac{1}{2}$. More importantly, this result is an exact one. So it is also valid even for the thermodynamic quantities are far from their critical values. Therefore, we have an analytical form for the order parameter $\Delta\Phi$.

\section{Summary}
\label{Summary}

In this paper, we have studied the equal area law and the thermodynamic criticality in the $Q$-$\Phi$ plane for the charged AdS black holes.

It is widely known that there exists the black hole phase transition for the charged AdS black holes. The phase transition point can be well determined by constructing the equal area law. However, different from the typical expression of the equal area law, we found that there is another expression, which is because that the charge $Q$ is a double-valued function of $\Phi$. When considering it, we obtained a new expression for the equal area law, see (\ref{eqa2}) or (\ref{jfs}). Based on these two different expressions of the equal area law, we obtained the analytical coexistence curve. Although these two expressions behave quite different, both of them confirm the same result, see (\ref{ww1}).

Making use the analytical coexistence curve, we explored the phase diagram and critical exponents in the $Q$-$\Phi$ plane. Different black hole phases are clearly displayed in Fig. \ref{TPhasd}. Near the critical points, we found that $(\Phi_{1}-\Phi_{\rm c})$ and $(\Phi_{3}-\Phi_{\rm c})$ have the same exponent of $\frac{1}{2}$. The order parameter $\Delta\Phi$ also has an exponent of $\frac{1}{2}$. More interestingly, the order parameter has a compact expression $\Delta\Phi=2(6\pi P)^{\frac{1}{4}} \times\left(Q_{\rm c}-Q\right)^{\frac{1}{2}}$ for all the range of the charge. Furthermore, we believe that our results are very useful on constructing the equal area law for a thermodynamic system when the double-valued function is included in.

\section*{Acknowledgements}
This work was supported by the National Natural Science Foundation of China (Grants No. 11675064). S.-W. Wei was also supported by the Chinese Scholarship Council (CSC) Scholarship (201806185016) to visit the University of Waterloo.


\end{document}